\documentclass{elsart}
\usepackage{epsfig,graphicx, amssymb}
\begin{document}
\runauthor{Litak et al.}

\begin{frontmatter}
\title{Stochastic description of the deterministic Ricker's population model.}   
\author[Lublin1]{Arkadiusz Syta},
\author[Lublin2]{Grzegorz Litak\thanksref{E-mail}}

\address[Lublin1]{Department of Applied Mathematics, Technical University
of
Lublin,
Nadbystrzycka 36, PL-20-618 Lublin, Poland}

\address[Lublin2]{Department of Applied Mechanics, Technical University of
Lublin,
Nadbystrzycka 36, PL-20-618 Lublin, Poland}

\thanks[E-mail]{Fax: +48-815250808; E-mail:
g.litak@pollub.pl (G. Litak)}

\begin{abstract}
We adopt the '0-1' test for chaos using
Brownian motion chains 
to identify the dynamics of the
Ricker's population model. In the '0-1' test '0' is related to 
regular  motion while '1' is associated with
chaotic 
motion.  The identified regular and chaotic types of solutions  have been confirmed 
by means of recurrence plots.
\end{abstract}

\end{frontmatter}

%As
%calculation of Lyapunov exponents meets some difficulties in case of
%some particular maps related to non-smooth dynamics. 

The Ricker's model \cite{Ricker1954} has been invented to analyze the population of 
fish. Originally the model used for stock assessment to learn and predict the population of 
fish
and to estimate how many fish can be caught
without impacting future production of fish. 
The dynamics of population change is given by the following map  \cite{Ricker1954}:
%eq1
\begin{equation}
x_{n+1}=F(x_n),~~~~ F(x_n)=Ax_{n}\exp(-x_n), 
\label{eq1}
\end{equation}
where $x_n$ means the population of consecutive fish generations and  the constant $\mu$ depends 
on biological conditions, $F(x_n)$ is a discrete transform function.
A similar model has been also used in the analysis of insect data by Moran \cite{Moran1950}.
This simple  model give complex answer for different system parameters.
It has attracted researchers interested in emerging the chaotic solutions in mathematical biology 
\cite{Murray1989}. Recently, it has been also analyzed in  presence of periodic or noisy forcing
\cite{Summers2000,Binder2005}.  
 
\begin{figure}[htb]
\centerline{
\epsfig{file=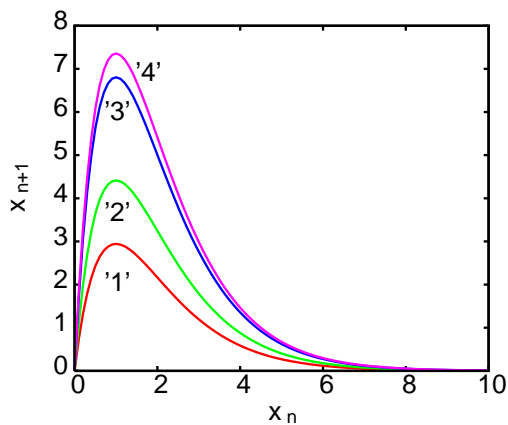,width=6.5cm,angle=-90}}
 \caption{\label{fig1}
Transform function $F(x_n)$ (Eq. \ref{eq1}). '1' denotes the curve 
for $\mu=8$ while '2'-'4'; $\mu=12$, 18.5 and 20, respectively. }
\end{figure}

In this paper we will apply Eq. (\ref{eq1}) to generate the time series of fish generations which can be
latter analyzed by nonlinear stochastic methods.  
 The corresponding transform functions $F(x_n)$ for  chosen values  of the control parameter $\mu$ ($\mu=8$, 12 
18.5 and 20) are 
plotted 
in Fig. \ref{fig1}.  

\begin{figure}[htb]
\centerline{
\epsfig{file=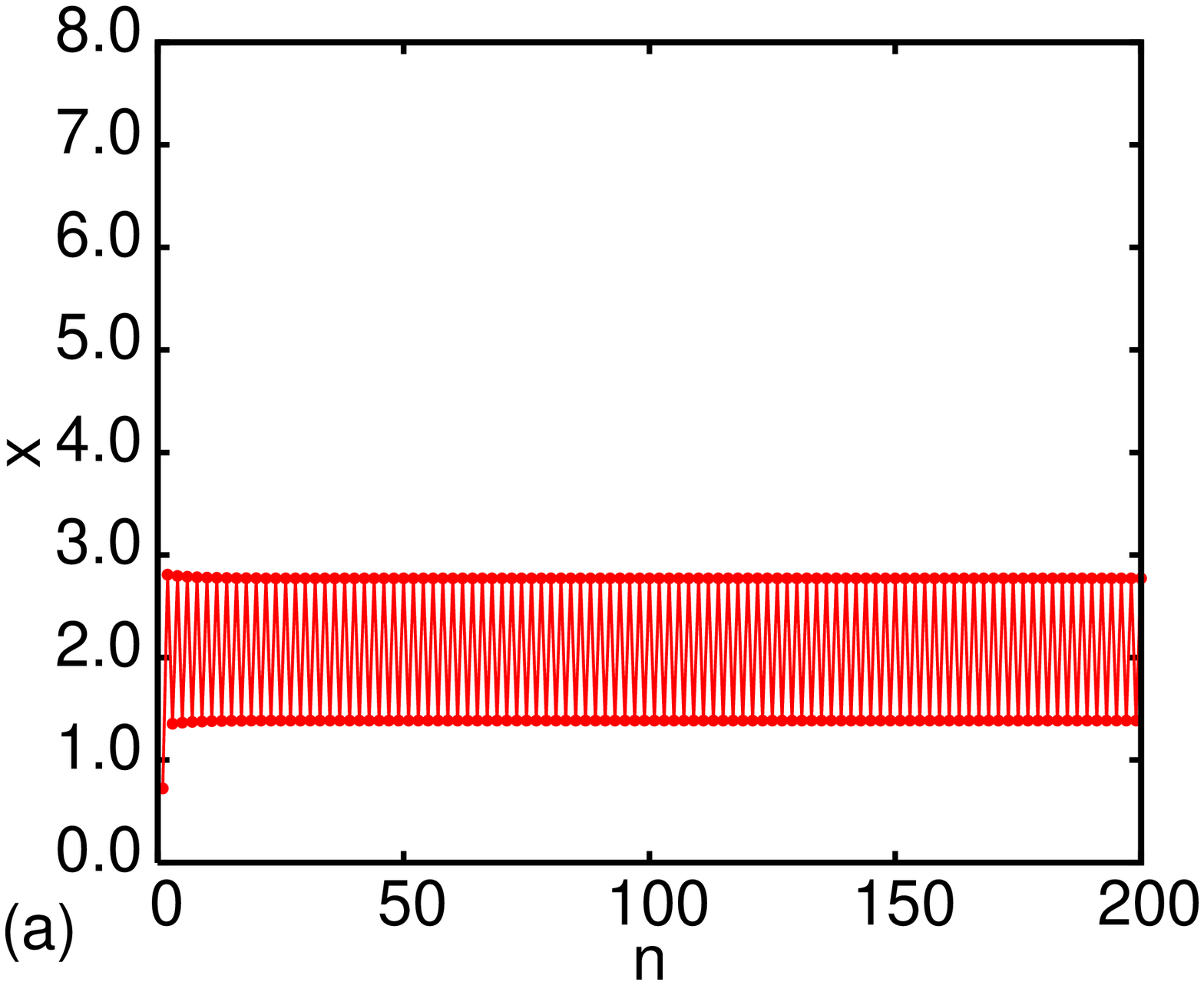,width=6.5cm,angle=-90}
\epsfig{file=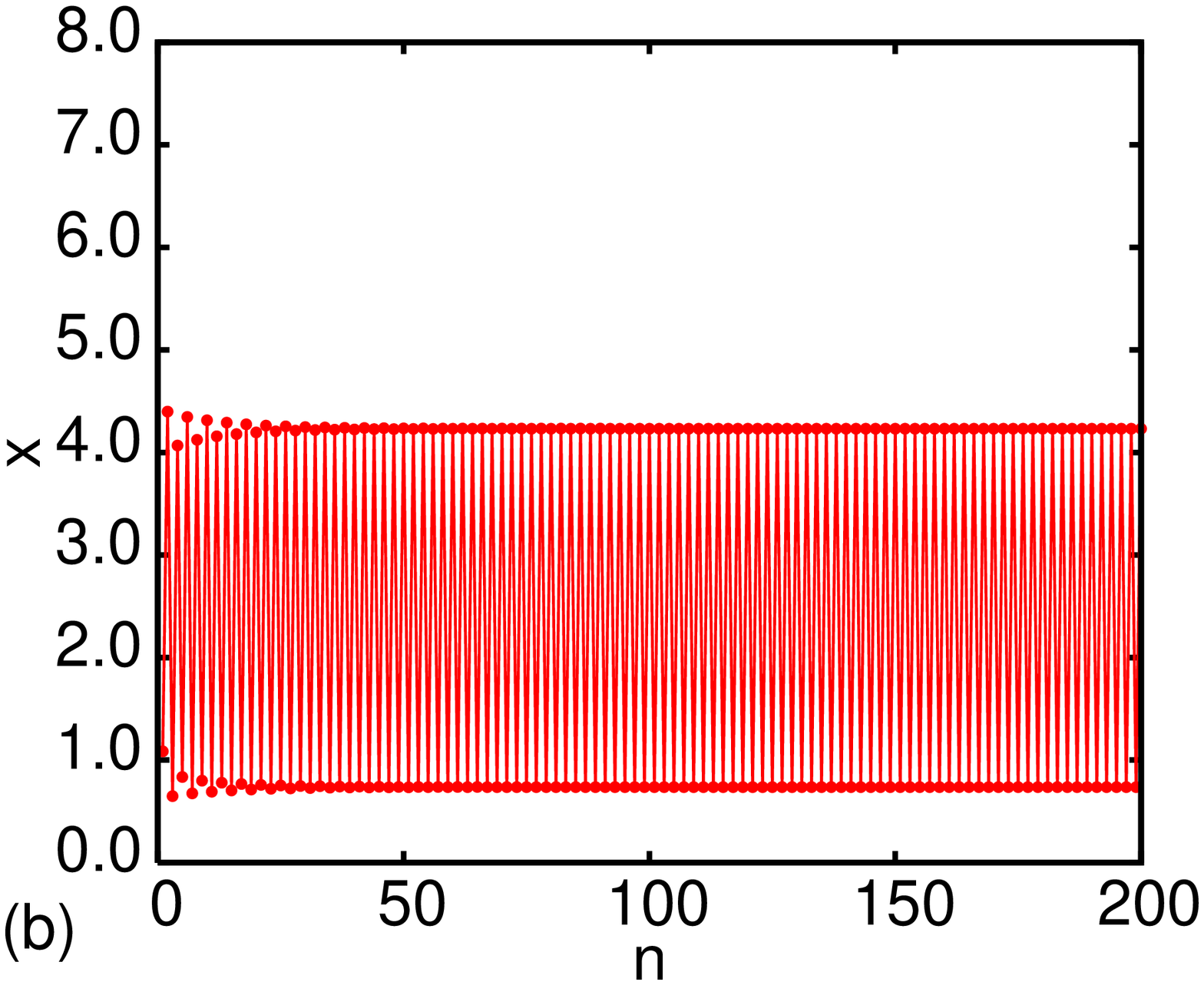,width=6.5cm,angle=-90}}

\centerline{
\epsfig{file=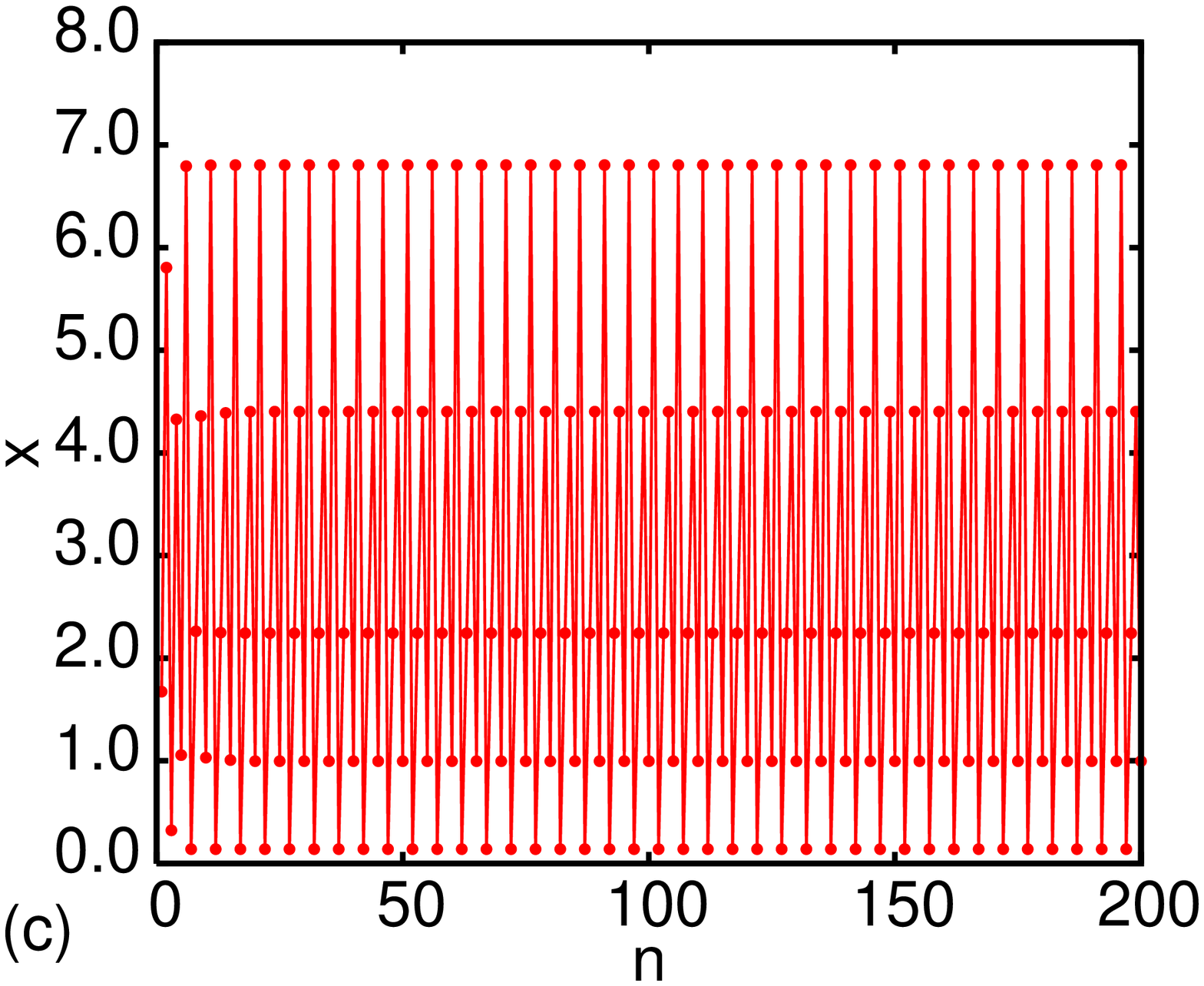,width=6.5cm,angle=-90}
\epsfig{file=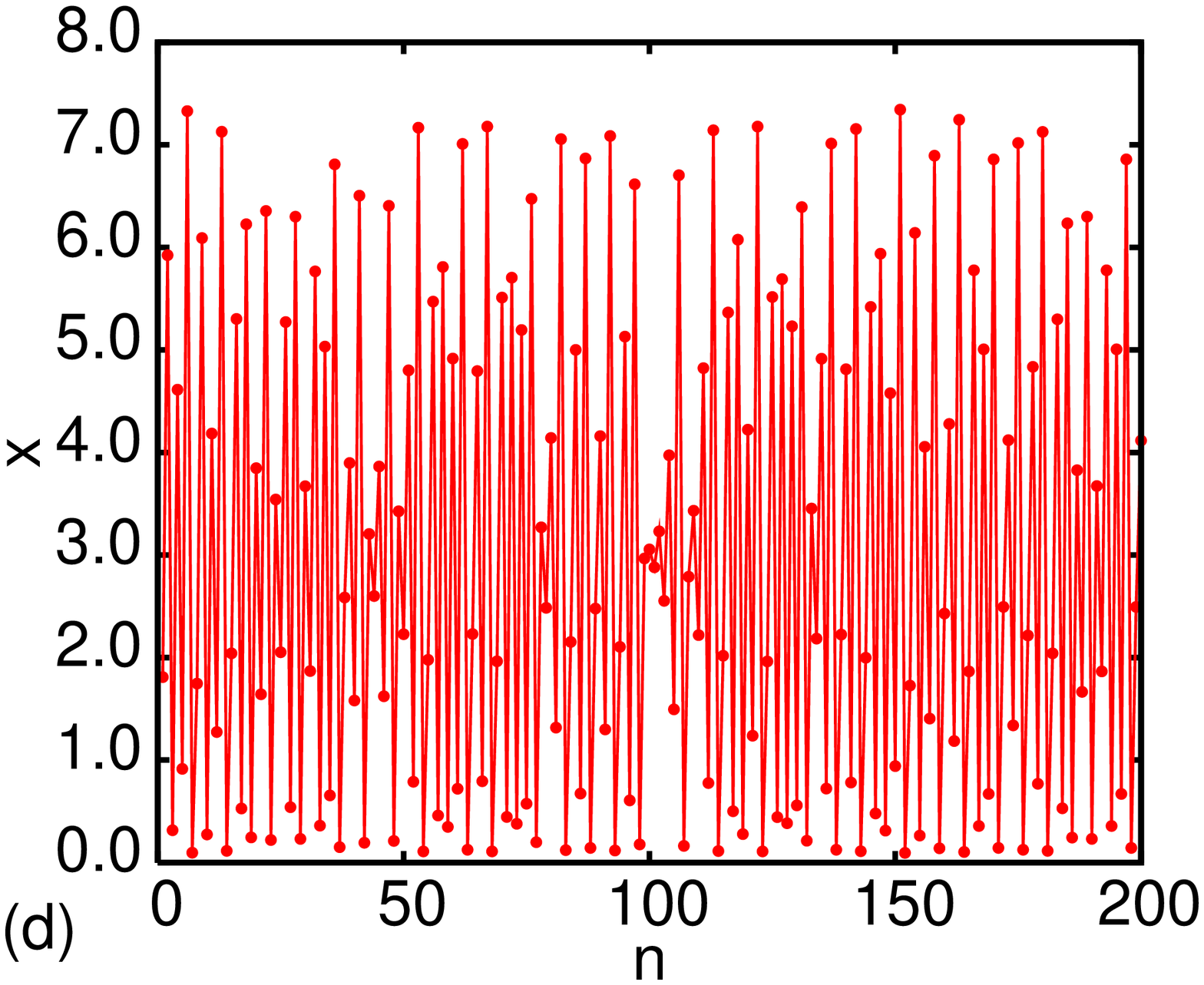,width=6.5cm,angle=-90}}
 \caption{
\label{fig2}
Time histories of $x_n$ (a-d) for $\mu=8$, 12, 18.5 and 20, respectively.  The initial point was set to $x_1=0.1$.}
\end{figure}

We started
examination of the dynamics of the above model with the time histories simulated for 
above chosen (Fig. \ref{fig1})  $\mu$ values. The time histories are 
presented in Fig. \ref{fig2}a-d. 
First three figures Figs. \ref{fig2}a-c correspond to periodic solutions with two (Figs. \ref{fig2}a-b) or four points
(Fig. \ref{fig2}c). The last figure (Fig.  \ref{fig2}d) represents a chaotic solution.  Note that  
the steady state for periodic solutions is reached 
after 
few initial iteration points.

\begin{figure}[htb]
\centerline{
\epsfig{file=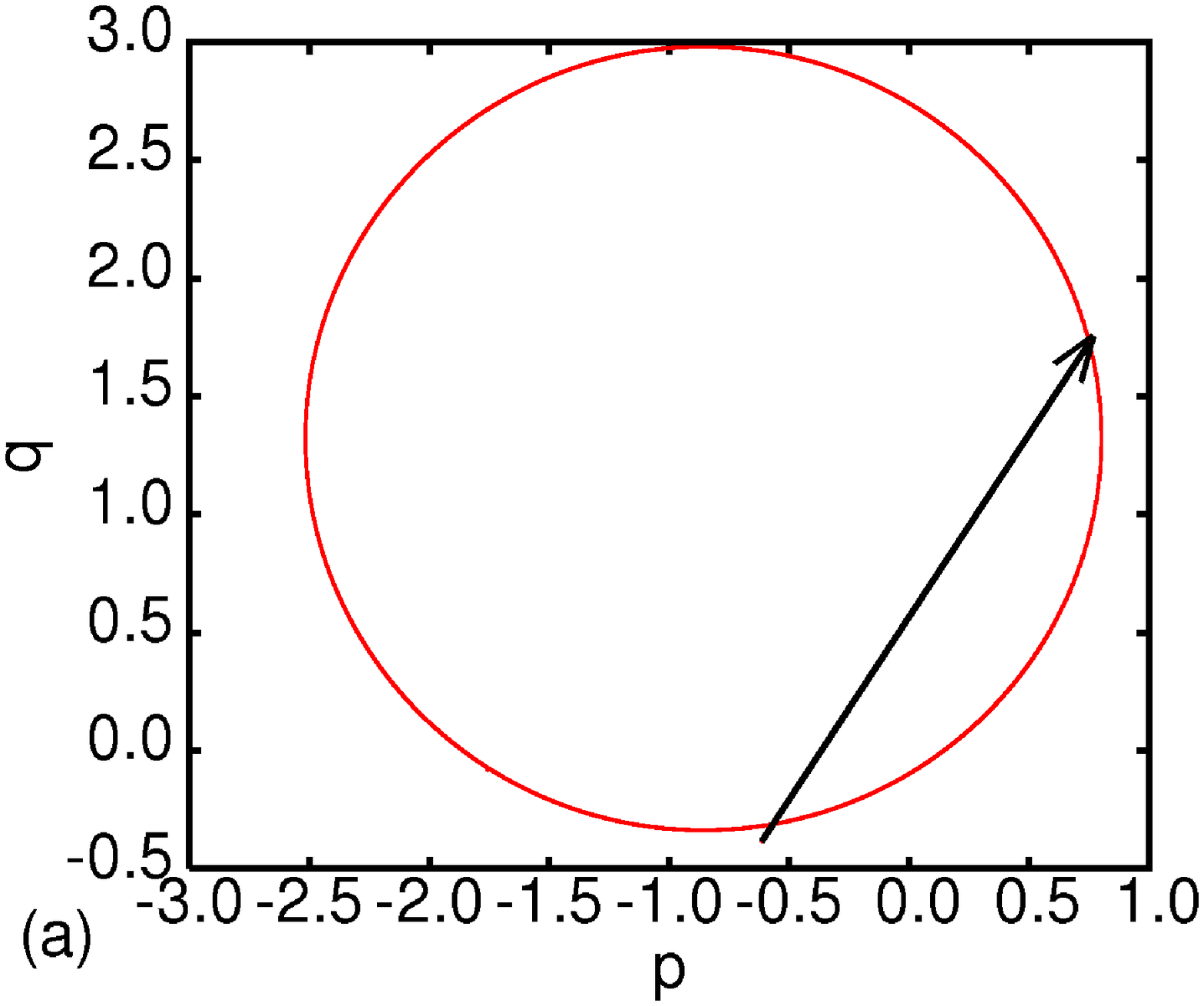,width=7.5cm,angle=0}
\epsfig{file=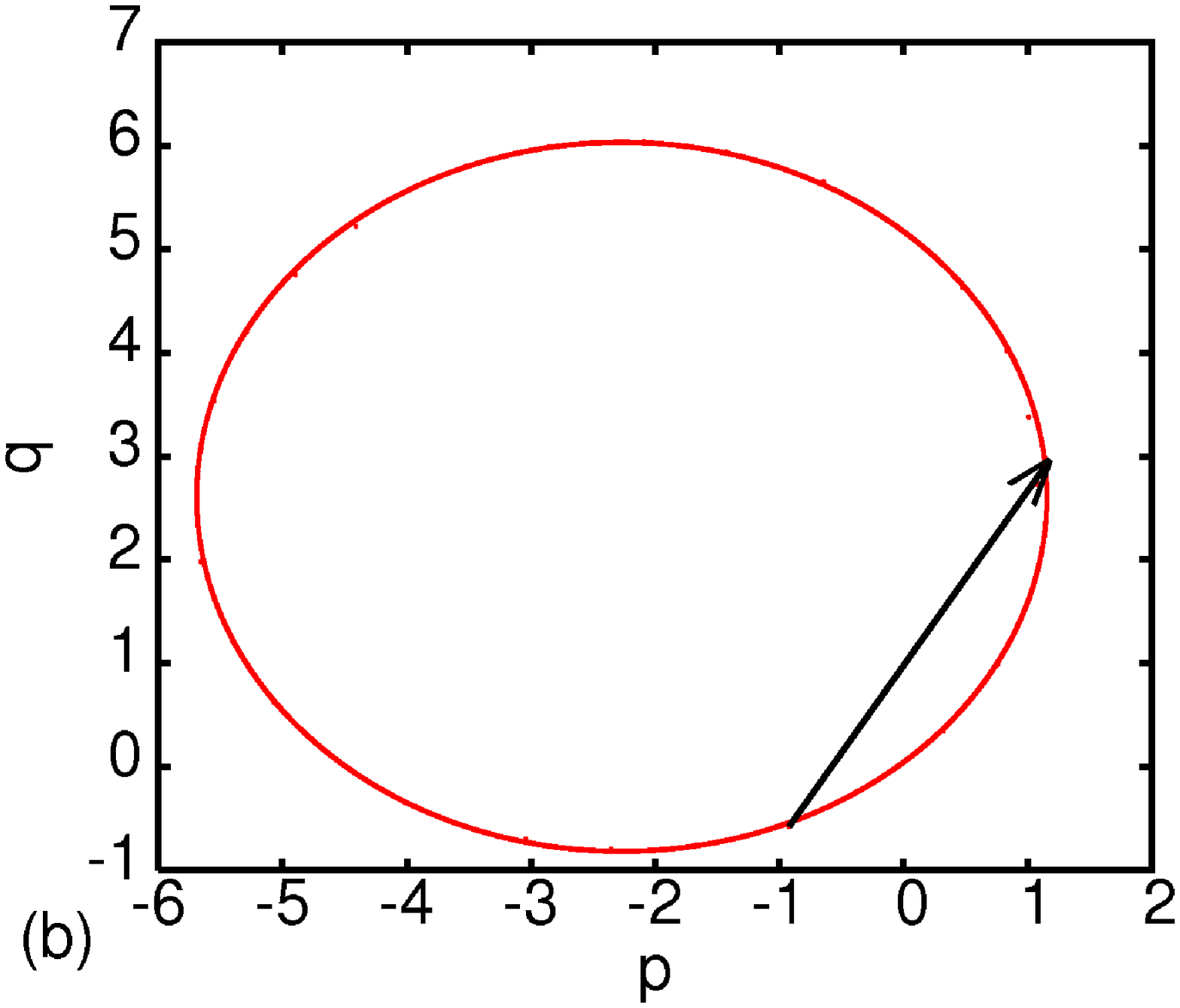,width=7.5cm,angle=0}}
\centerline{
\epsfig{file=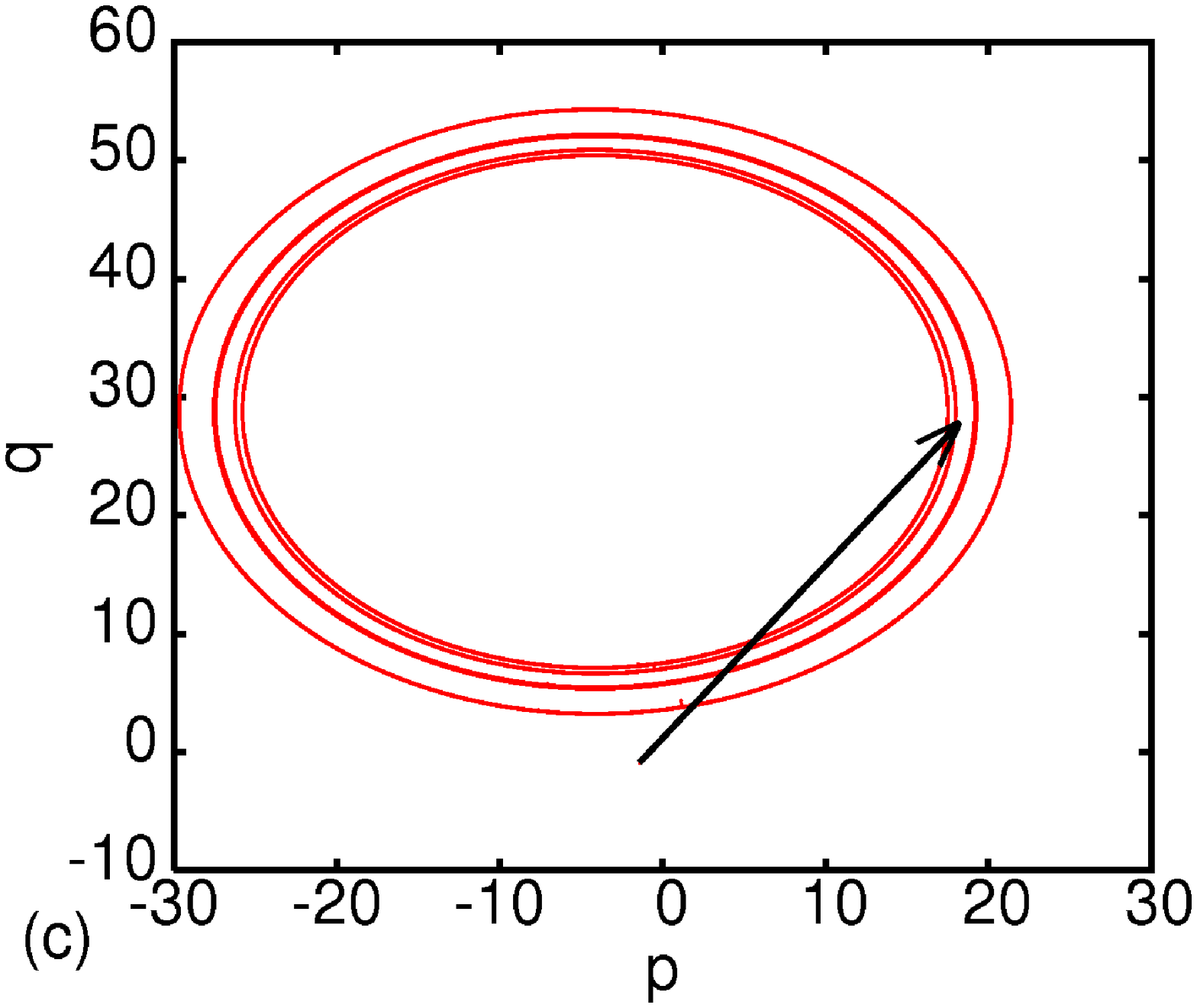,width=7.5cm,angle=0}
\epsfig{file=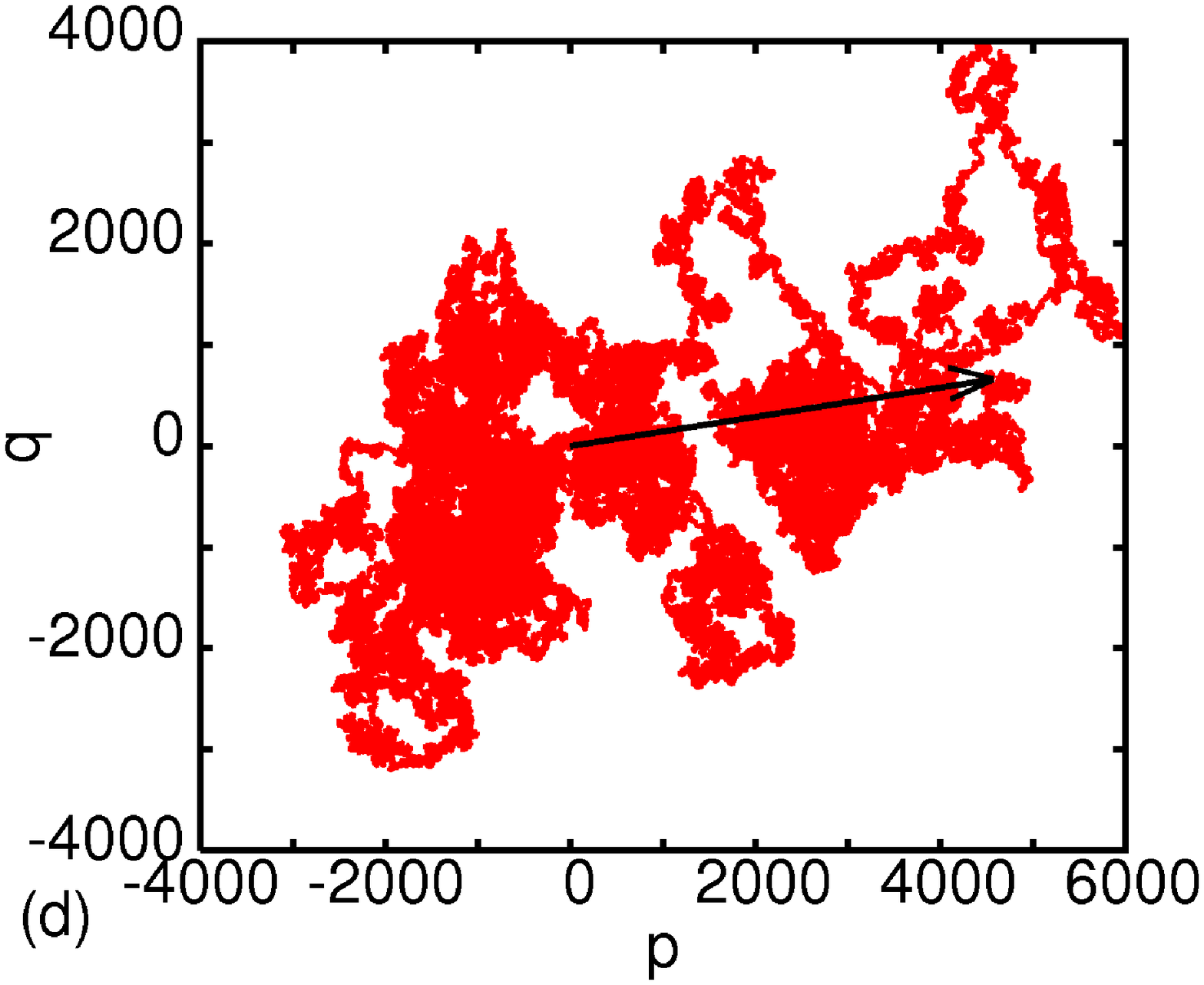,width=7.5cm,angle=0}}
 \caption{
\label{fig3}
'Phase portrait' in $p-q$ coordinates. Here $p(n)$,$q(n)$ have been calculated for  $n=1,...,4000000$. Arrows show the total 
displacement from the first to last 
points. The first point was set by assuming $x_1=0.1$. Note the scale in each figure is different. }
\end{figure}

\begin{figure}[htb]
\centerline{
\epsfig{file=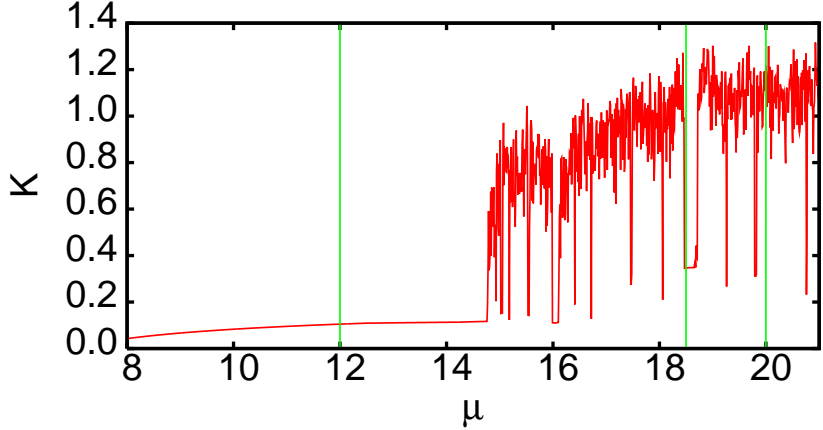,width=6.5cm,angle=-90}
}
 \caption{\label{fig4} $K$ versus $\mu$. Vertical straight lines correspond to  $\mu=8$, 12, 18.5 and 20 respectively. }
\end{figure}

%fig5
\begin{figure}[htb]
\centerline{
\epsfig{file=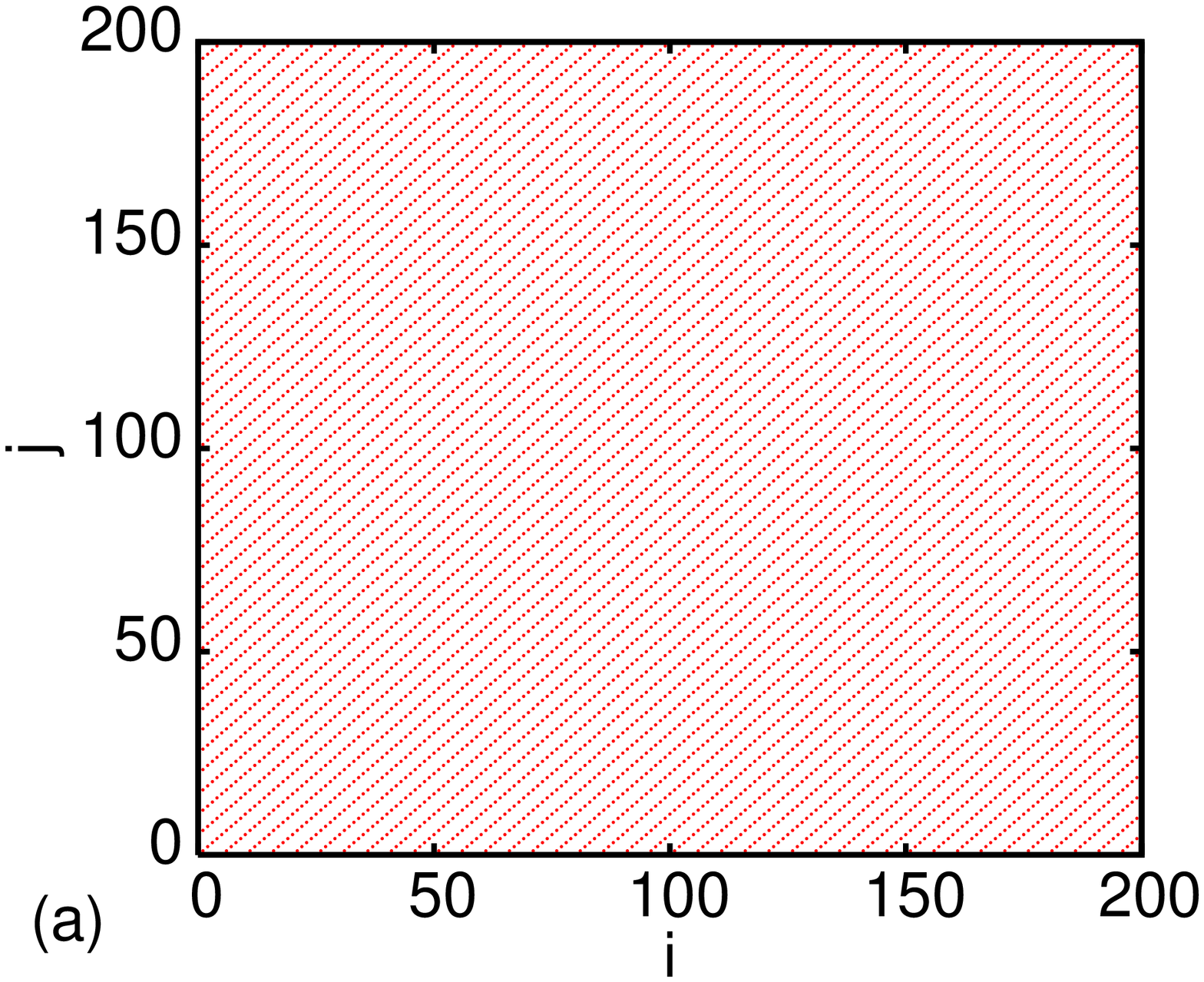,width=8.5cm,angle=-90}}
\centerline{
\epsfig{file=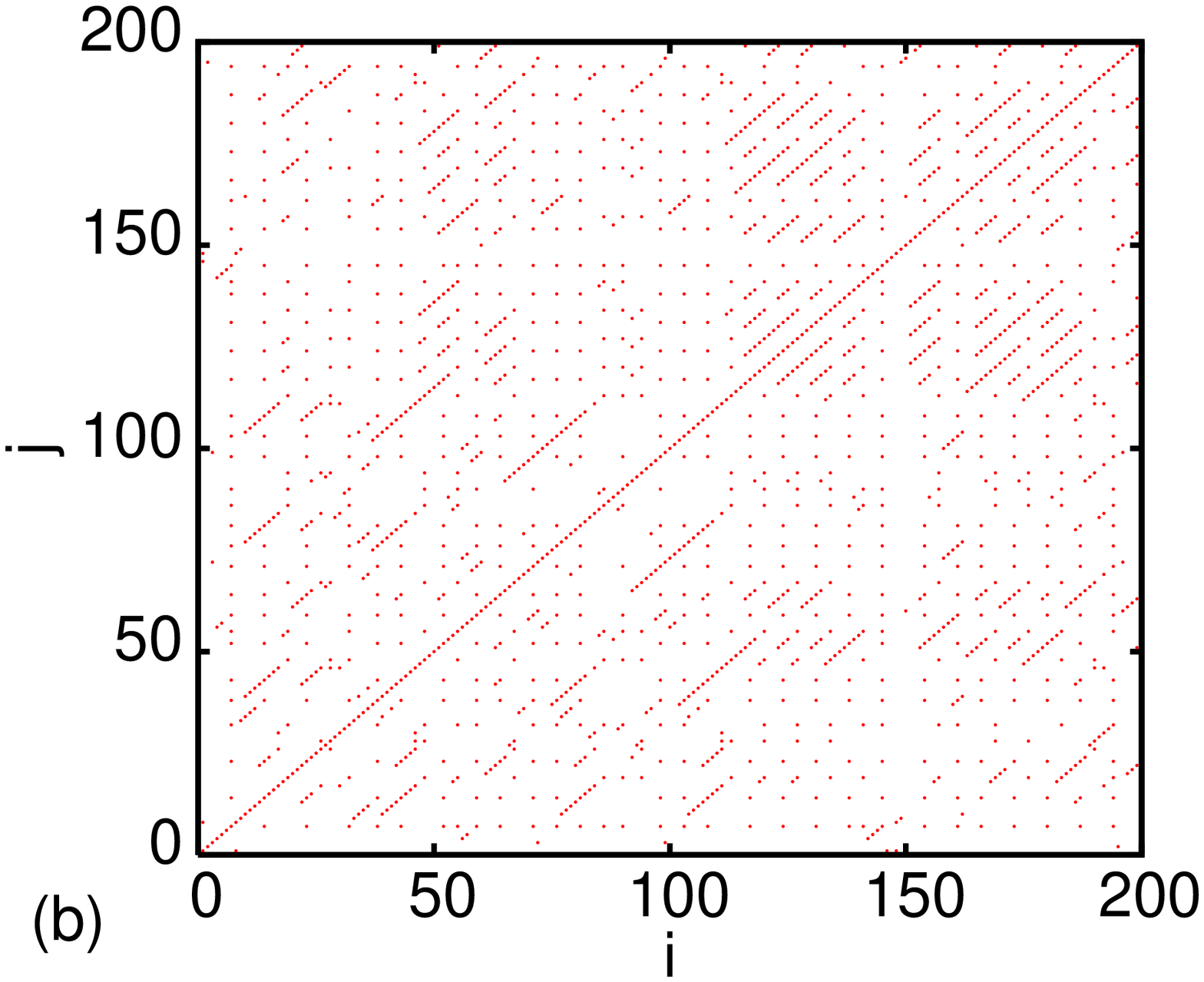,width=8.5cm,angle=-90}}
 \caption{ \label{fig5} Recurrence plots of $x_n$ (a-b) for $\mu=18.5$ and 20 respectively.}
\end{figure}

Naturally the standard way 
to classify the dynamics of   
nonlinear time series is 
to perform
reconstruction of phase space \cite{Takens1981} and calculate the Lyapunov exponent \cite{Wolf1985,Kantz1994}. Alternatively 
one can
 adopt of '0-1' test
method introduced by Gottwald and Melbourne  \cite{Gottwald2004}. The method was
previously successfully tested for the Van
der Pol
system, the Kortweg-de Vries equations \cite{Gottwald2004} the logistic map, and the Lorenz 96 system, including the system with  weak noise 
\cite{Gottwald2005}.

Let us  rearrange the  $x_n$ series in the following way: 
%eq2
\begin{equation}
p(n) = \sum_{j=1}^n x_j \cos(jc),
\label{eq2}
\end{equation}
for consecutive $n=1$, 2, 3, ... .
Here the constant $c=3.7$ has been chosen arbitrary as in Ref. \cite{Gottwald2004,Gottwald2005}.

Note that we build new series of $p(n)$ (Eq. \ref{eq4})
starting with an essentially bounded coordinate $x_n$. 
However, the new series can be either 
bounded or unbounded depending on dynamics of the examined process: regular or chaotic, respectively.
In fact for a random noise $\Gamma_n$ we have  a Brownian motion analogy \cite{Risken1989,Gaspard1998,Cecconi2005}:
\begin{equation}
p(n+1) =p(n)+ \Gamma_n~~~n=1,2,...,N
\end{equation} 
Here $p(n)$ represents unbounded time series where total displacement is scaled by the square root of  steps number $N$ 
\cite{Risken1989}.  $\Gamma_n$ is a diffusion term which couples different states $n \neq m$.    
In our case 
this term is given by the formula \cite{Gottwald2004,Gottwald2005}:
%eq4
\begin{equation} 
\Gamma_n=x_n \cos(nc).
\label{eq4}
\end{equation}
Needless to say, in case of chaotic time series $x_n$,  $\Gamma_n$ will mimic random behaviour 
\cite{Gaspard1998,Cecconi2005}
leading to an unbounded displacement. 

On the other hand
for 
a periodic solution $x_n$:
%eq5
\begin{equation}
\{x_n\} \rightarrow x_1,...,x_r,x_1,...,x_r, x_1,...,x_s, ~~~~~{\rm for}~~~s < r,
\label{eq5}
\end{equation}
where the index $r$ is related to a $r$-element period included in $x_n$ series. If it appears $j$ times in the total time series $x_n$ then  
$n=rj+s$. 
The parameter 
$p(n)$ can be expressed  
as (see Appendix A):
%eq6
\begin{equation}
p(n)=p(jr+s)= \sum_{j=1}^n x_j \cos(jc) \approx  
 -\sum_{k=1}^{r+ \Theta(s-k)} x_k R_{nr} \sin (kc + \phi(n_r)),  \label{eq6}
\end{equation}
where $\phi(n_r)$ and $R_{nr}$ is the angle a characteristic circle radius (Eq. \ref{eqA.11}, here $R_{nr} \approx 1$) dependent on the 
number of repetitions of a period $x_1,...,x_r$ and a value $c$ (Appendix A)
%eq7 
and $\Theta(\cdot)$ is the Heaviside step function. 

Defining the other complementary quantity $q(n)$ in the similar way:
%eq8
\begin{equation}
\label{eq8}
q(n) = \sum_{j=1}^n x_j \sin(jc)
\end{equation}
we can derive a corresponding simplified expression for a periodic series $x_n$ (Eq. \ref{eq5}) in an analogous way.
Thus an appropriate  expression (Appendix A) reads: 
\begin{equation}
\label{eq9}
q(n)=q(jr+s)= \sum_{j=1}^n x_j \cos(jc) \approx
 \sum_{i=1}^{r+ \Theta(s-i)} x_i  R_{nr} \cos (ic + \phi).
\end{equation}
Note the pair ($q,p$) defines new coordinates in the two dimensional space.
Knowing that $r$ and $s$ (Eqs. \ref{eq6}, \ref{eq9}) are usually low natural numbers (see  Fig. \ref{fig2} $r=2$ or 4) we can 
conclude that
one can  easily see  if series  $p(n)$ and $q(n)$ are bounded for arbitrary large $n$.
Furthermore the simple expressions Eqs. (\ref{eq6}) and (\ref{eq9}) determine their trajectory in a  circle like shape
which  radius $R$ can be easily estimated as 
%eq10
\begin{equation}
\label{eq10}
R \approx r <x_i>. 
\end{equation}
In the above expression $<\cdot>$ denotes the average value while $r$ is a period of $\{x_n\}$ (Eq. \ref{eq5}). 
 In Fig. \ref{fig3} we show examples of trajectories (plotted by consecutive points) for the examined system with four values of 
A as in Figs. 
\ref{fig1}-\ref{fig2} calculated from $x_n$ series using Eqs. (\ref{eq2},\ref{eq8}) and  for  $n=1,...,4000000$. 
 Arrows show the total displacement from the first ($n=1$) to last points ($n=4000000$). 
Thus depending on motion type the transformed system, into $p$ and/or $q$  coordinates, their time series show the 
bounded or unbounded
behaviour.

To classify the dynamics in a more systematic way,
we have performed the next step by
calculation mean square displacement of $p(n)$ \cite{Gottwald2004,Gottwald2005}: 

%eq11
\begin{equation}
M(n)= \lim_{N \rightarrow \infty} \frac{1}{N} \sum_{j=1}^{N} \left(p(j+ n) - 
p(j)\right)^2
\label{eq11}
 \end{equation}
together  its asymptotic behaviour of $M(n)$ estimated by the following limit \cite{Gottwald2004,Gottwald2005}:
%eq12
\begin{equation}
K= \lim_{n \rightarrow \infty} \frac{\ln (M(n)+1)}{\ln n} 
\label{eq12}
\end{equation}
Although the limits in the above two expressions go to $\infty$, in practice, it is enough to assume 
%eq13
\begin{equation}
N=150000~~~ {\rm and}~~~ n=3000000
\label{eq13}
\end{equation}
 to get a clear distinction between regular and 
chaotic states.
The results for $K$ versus $\mu$ are plotted in Fig. \ref{fig4}. 
It is apparent that in case of chaotic oscillations  $K \approx 1$ (mainly above $\mu=14.6$)
while for regular motion $K \approx 0$ (see region below $\mu=14.6$). Some variations around 0 and 1 can be 
explained by a finite truncation (Eq. \ref{eq13}). As noted by Gottwald and Melbourne
\cite{Gottwald2005}
the choice 
of the parameter $c$ can be sometimes important 
creating resonances which should be avoided. However
we have checked this possibility in our calculations and assumed it 
eliminates such unwanted cases.
Note that there exists a sudden change of  $K \approx 1$
into lower value $K \approx 0.38$ for $\mu$ about $18.5$.  Looking at Fig. \ref{fig2}c  we know that the system 
is  periodic
 with a characteristic period $r=4$.
It is also bounded in terms of coordinates $p(n)$ and $q(n)$ (as can be seen in Fig. \ref{fig3}c)  but the characteristic 
circle radius 
$R$
in ($p,q$) space is
relatively large (Fig. \ref{fig3}c, Eq. \ref{eq10}) comparing to other regular states (Fig. \ref{fig3}a-b).  It is clear from 
Eq. \ref{eq10} as  $r$ is relatively large here. 
To show the difference between both cases $\mu=18.5$ and 20 we plotted corresponding recurrence plots defined by 
the scalars $p(n)$ (without embedding) \cite{Thiel2004}.  Naturally the usual way is to investigate a recurrence 
plot in a reconstructed phase space \cite{Webber1994}. But as has been recently  shown by  Thiel {\em at al.} 
\cite{Thiel2004} 
treatment without embedding  gives correct estimation of
dynamical invariants and various recurrence plot analysis measures can be applied \cite{Marwan2004}.
Here we plot a dot 
an a squared matrix whose axes correspond to the $i$ and $j$ indices
if only the condition 
\begin{equation}
\|x_i - x_j \| < \epsilon
\end{equation}
is full filed, where  $\epsilon$ is a certain threshold value.
Using this method we can again classify independently the states of the system as 
regular  for $\mu=18.5$ (Fig. \ref{fig5}a) with regular patterns (continuous diagonal lines) and
chaotic for 
$\mu=20$, where the lines patterns appear with different finite length distributed in some random way.
It indicates that the system is visiting the same region of the attractor many times \cite{Webber1994}.
The length of these short diagonal lines can be related (inversely proportional) to the largest positive  Lyapunov exponent $\lambda_1$ 
\cite{Eckmann1987,Webber1994}. 

In summary, the 0-1 test for chaos appeared to be reliable method in case of 
Ricker's model. Using a certain projection of the original time series $\{x_n\}$  into $p(n)$ we could use the stochastic 
methods to analyze the asymptotic behaviour of mean square displacement (in $p(n)$) .
It enabled us  
to classify the system dynamics directly from scalar signal 
without any procedure of 
phase space 
reconstruction. 
Like in other systems analyzed by Gottwald and Melbourne \cite{Gottwald2004,Gottwald2005} 
for chaotic changes of consecutive fish populations we observed clear oscillations around $K =1$.   
On the other hand  in case of  regular systems it is more useful if the period of repeating chain in a time series 
is shorter.
However in longer periods ($r=4$ for $\mu =18.5$ Fig. \ref{fig4}) we observed $K \approx 0.38$. We hope that
this disadvantage can be reduced in future by using the estimated radius $R$ (Eq. \ref{eq10}) in the definition of 
$K$. The above method seams to be very useful for analysis of experimental data
as well as  for simulated dynamical systems especially those having non-continuous or non-smooth vector fields 
\cite{Wiercigroch1997,Litak2002,Borowiec2006}.

{\bf Acknowledgements:} 
GL would like to thank Max Planck Institute for the
Physics of Complex Systems in Dresden for hospitality.

\appendix{\Large \bf \noindent Appendix A}
\def\thesection{A}
\setcounter{equation}{0}
\def\theequation{A.\arabic{equation}}  % dla stylu 'article'

For a time series represented as a large multiple sequence $x_1,...,x_r$

%eqA.1
\begin{equation}
x_1,...,x_r,x_1,...,x_r,x_1,...,x_s
\label{eqA.1}
\end{equation}

%eqA.2
\begin{equation}
 \sum_{j=1}^n x_j \cos(jc) = \sum_{i=1}^{s} x_i \left( \sum_{k=1}^{n_r} \cos(ic+kc) \right)
                            +\sum_{i=s}^{r} x_i \left( \sum_{k=1}^{n_r} \cos(ic+kc) \right),
\label{eqA.2}
\end{equation}
where $n_r$ corresponds to the multiplicity of a shorter $r$-element ($x_1,...,x_r$) period inside $x_n$ time series. 
Using a simple decoupling
%eqA.3
\begin{equation} 
\cos(ic + kc)= \cos(ic)\cos(kc)-\sin(ic)\sin(kc)
\label{eqA.3}
\end{equation}
and
%eqA.4
\begin{eqnarray}
&& A_{nr}=\sum_{k=1}^{n_r} \cos(kc) \label{eqA.4} \\
&& = \frac{ \cos c -\cos c (n_r +1)-1 +\cos c \cos c(n_r +1) + \sin c \sin c(n_r +1)}{2 - 2 \cos c}.
\nonumber
\end{eqnarray}
Similarly
%eqA.5
\begin{eqnarray}
&& B_{nr} =\sum_{k=1}^{n_r} 
\sin (kc) \label{eqA.5} \\ 
&& =\frac{ -\sin c \cos  c(n_r +1)    +\cos c \sin  c(n_r +1) +\sin c}{
2 - 2 \cos c}
\nonumber
\end{eqnarray}
Figure A.1 presents results $B_{nr}$ versus $A_{nr}$ for two different values of $c$ ($c=1.7$ and $c=3.7$)
note that in both cases the points are lying on circles. They are however centered in different points and have
different radii $R_c$.  In case of $c=3.7$ the radius is less than 1.   
The centers are given by the averages taken from Eq. \ref{eqA.4} and \ref{eqA.5}.
%eqA.6
\begin{equation}
A_{nr0}=0.5 {\rm~~~for~ any~~~}  c,
%eqA.6
\end{equation}
 while 
%eqA.7
\begin{equation}
B_{nr0}= \frac{\sin c}{2- 2 \cos c}.  
%eqA.7
\end{equation}
Note, using the circles representation, both expressions for $A_{nr}$ and $B_{nr}$ (Eqs. \ref{eqA.4} and \ref{eqA.5}, 
respectively) can be 
now written in a 
compact form
%eqA.8
\begin{eqnarray}
&& A_{nr}= A_{nr0} + R_c \cos \psi(n_r), \nonumber \\
&& B_{nr}= B_{nr0} + R_c \sin \psi(n_r).
\label{eqA.8}
\end{eqnarray}
Interestingly for $c=3.7$ we have found numerically  a change of  $\psi$ versus $n_r$ is linear 
\begin{equation}
\psi(n_r) \approx 0.5584 n_r.
\end{equation}
Using the above notation we can write 
%eqA.7
\begin{eqnarray}
&& \sum_{j=1}^n x_j \cos(jc)= \sum_{k=1}^{r+ \Theta(s-k)} x_k \left(A_{nr} \cos (ck)  - B_{rn}\sin 
(kc)\right) \nonumber \\
&& =  \sum_{k=1}^{r+ \Theta(s-k)} x_k R_{nr} \cos (ck+ \phi (n_r)) 
\end{eqnarray} 
where 
%eqA.8
\begin{equation}
\tan (\phi)= \frac{A_{nr}}{B_{nr}},~~ R_{nr}=\sqrt{A_{nr}^2+B_{nr}^2},
\label{eqA.8}
\end{equation}
where $\Theta(\cdot)$ is a Heaviside step function.

\begin{figure}
\centerline{
\epsfig{file=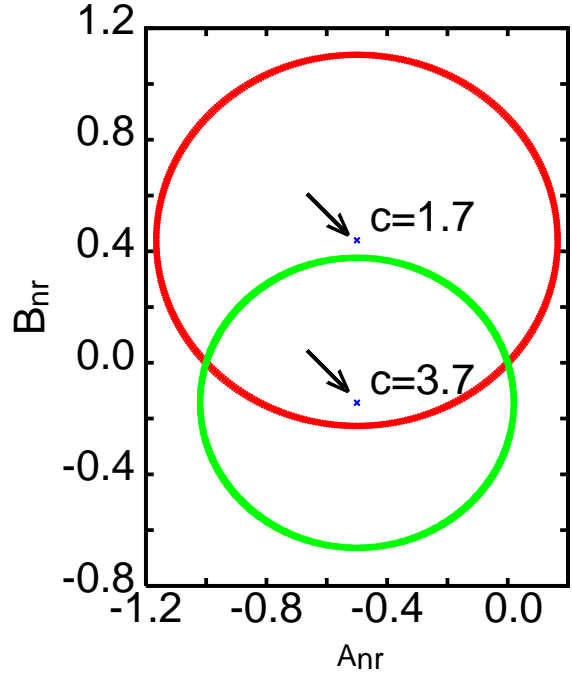,width=10.5cm,angle=-90}}
 \caption{ \label{figA.1} $B_{nr}$ versus $A_{nr}$ for $c=1.7$ and $c=3.7$, respectively. Arrows indicate corresponding
circle centers.}
\end{figure}

Similar analysis can be applied for
%eqA.12
\begin{equation}
 \sum_{j=1}^n x_j \sin(jc)
= \sum_{i=1}^{s} x_i \left( \sum_{k=1}^{n_r} \sin(ic+kc) \right)
                            +\sum_{i=s}^{r} x_i \left( \sum_{k=1}^{n_r} \sin(ic+kc) \right),
\label{eqA.12}
\end{equation}
where

%eqA.13
\begin{equation}
\sin(ic + kc)= \sin(ic)\cos(kc)+\cos(ic)\sin(kc).
\label{eqA.13}
\end{equation}
Therefore
%eqA.14
\begin{eqnarray}
&& \sum_{j=1}^n x_j \cos(jc)= \sum_{k=1}^{r+ \Theta(s-k)} x_k \left(A_{nr} \sin (ck)  + B_{rn}\sin
(kc)\right) \nonumber \\
&& =  \sum_{k=1}^{r+ \Theta(s-k)} x_k R_{nr} \sin (ck+ \phi (n_r)).
\label{eqA.14}
\end{eqnarray}

\end{document}